\documentclass[12pt,a4paper]{article}
\usepackage[truedimen,margin=30mm]{geometry} 

\usepackage{mathrsfs}
\usepackage{amssymb}
\usepackage{amsmath}
\usepackage{ascmac}
\usepackage{amsthm}
\usepackage[dvips]{graphicx}
\usepackage{natbib}
\usepackage{setspace}
\usepackage{times}
\usepackage{color}

\usepackage{titlesec}
\titleformat*{\section}{\large\bfseries}
\titleformat*{\subsection}{\it}

\newtheorem{thm}{Theorem}

\newtheorem{cor}{Corollary}

\def\ep{{\varepsilon}}

\title{{\bf Beyond Tweedie's Formula: Conditional Score Modeling for Empirical Bayes Inference}\footnote{\today}}

\date{}

\begin{document}

\maketitle
\doublespacing

\vspace{-1.5cm}
\begin{center}
{\large Shonosuke Sugasawa$^1$ and Zhigen Zhao$^2$}

\medskip
\noindent
$^1$Faculty of Economics, Keio University, Tokyo, Japan\\
$^2$Department of Statistics, Operations, and Data Science, Temple University, Philadelphia PA, USA
\end{center}

\vspace{2mm}
\begin{center}
{\bf \large Abstract}
\end{center}
We propose conditional f-modeling (Cf-modeling), a framework for empirical Bayes inference with covariates. A central identity shows that the conditional marginal score function determines not only the posterior mean through Tweedie’s formula, but also the posterior moment-generating function, providing a basis for recovering posterior quantities without explicit prior modeling. Motivated by this observation, we treat the conditional marginal score as the primary object of inference and estimate it directly using an energy-based representation and Hyv\"arinen score matching, thereby avoiding potentially intractable covariate-dependent normalizing constants. The resulting framework flexibly accommodates covariate effects and heteroscedasticity and provides a practical approach to posterior moment estimation and uncertainty quantification. We demonstrate the effectiveness of the proposed method through simulations and an RNA-seq application.

\bigskip\noindent
{\bf Key words}: Posterior distribution; Shrinkage estimation; Score matching; Unequal variance

\section{Introduction}\label{sec:intro}
Empirical Bayes methods provide a powerful framework for simultaneous inference in large-scale problems by borrowing information across many related units. Classical developments go back to the shrinkage estimation theory of \citet{stein1981estimation}, \citet{efron1973stein}, and \citet{morris1983parametric}. In the normal means problem, one observes noisy estimates of latent signals, and the goal is to estimate each signal by combining the individual observation with empirical distributional information in the full sample. This idea has been influential in high-dimensional inference, including multiple testing, genomics, small area estimation, and large-scale prediction \citep{efron2012large, efron2011tweedie, brown2009nonparametric, jiang2009general, stephens2017false, ignatiadis2022confidence}.

There are two different approaches in the existing literature, f-modeling and g-modeling \citep{efron2011tweedie, efron2016empirical}. In $g$-modeling, one specifies or estimates the prior distribution of the latent signal and then computes posterior quantities through Bayes' rule. This approach includes parametric normal-normal models, nonparametric maximum likelihood estimation of the mixing distribution, and convex optimization formulations \citep{kiefer1956consistency, laird1978nonparametric, jiang2009general, koenker2014convex, gu2017unobserved, saha2020nonparametric}. In contrast, $f$-modeling estimates the marginal density of the noisy observation directly. Through Tweedie's formula, posterior mean estimation can be reduced to estimating the score function of the marginal density, rather than the prior distribution itself. This observation has motivated empirical Bayes methods based on kernel density estimation, deconvolution, nonparametric likelihood, and convex optimization \citep{brown2009nonparametric, efron2016empirical, greenshtein2009application, koenker2014convex, narasimhan2020deconvolveR, efron2019bayes, ghosh2025stein}.

Despite these developments, existing $f$-modeling methods have two major limitations. 
First, classical $f$-modeling is primarily a framework for point estimation. 
Through Tweedie's formula, posterior mean can be obtained via score estimation, but full posterior inference, such as uncertainty quantification, confidence interval construction, and evaluation of posterior probabilities for events of interest, is not readily available \citep{efron2024empirical}. 
Second, traditional f-modeling approaches have largely focused on settings with marginally independent and identically distributed observations \citep{louis2019comment}. 
In many applications, each unit is accompanied by auxiliary information, and the distribution of the latent signal may vary systematically with this information. 
Examples include gene-expression analysis, small area estimation, and large-scale experimentation, where effect sizes or area-level parameters may depend on baseline expression levels, demographic variables, or pre-treatment characteristics \citep{fay1979estimates, rao2015small, love2014moderated, himes2014rna, ignatiadis2019covariate}. In such cases, empirical Bayes estimation requires the conditional marginal density, or more precisely its score, rather than a common marginal density shared by all observations. When the sampling variances are unequal, the marginal distribution also depends on the unit-specific variance, so observations are no longer exchangeable through a common marginal density. This makes direct conditional density estimation infeasible. 

This paper addresses both limitations simultaneously. When covariate information is available, we establish a posterior moment generating function identity showing that the conditional score function fully characterizes the posterior distribution. Under mild regularity conditions, a corresponding characteristic-function representation is also available. These results significantly extend the scope of $f$-modeling beyond posterior mean estimation, demonstrating that the conditional score function contains sufficient information to recover the entire posterior distribution. We term this framework {\it conditional $f$-modeling (C$f$-modeling)}. Unlike classical $f$-modeling, Cf-modeling naturally incorporates covariate information and remains applicable in settings with marginally non-identically distributed observations, greatly expanding the range of empirical Bayes problems that can be addressed within the $f$-modeling paradigm.

To estimate the conditional score, we minimize the Hyv\"arinen score, or equivalently the Fisher divergence between the true and modeled conditional marginal distributions \citep{hyvarinen2005estimation}. This criterion is particularly appealing in the empirical Bayes setting because the discrepancy between the true and estimated conditional scores directly determines the excess Bayes risk of the resulting Tweedie-type estimator \citep{ghosh2025stein}. 
Rather than focusing on score estimation in the classical normal means model, we develop a general framework for covariate-assisted empirical Bayes inference with unequal variances. The proposed framework establishes conditional Tweedie identities, accommodates flexible nonlinear covariate effects, and shows how the fitted energy function can recover posterior moments and uncertainty measures, thereby extending empirical Bayes estimation from point estimation to full posterior inference.

The proposed approach has three main features. First, unlike g-modeling approaches that seek to recover the latent mixing distribution, our approach operates directly on the conditional marginal score, avoiding the associated inverse problem while flexibly accommodating nonlinear covariate effects and unequal variances without specifying a parametric prior distribution. Second, estimation is simple: under a linear basis expansion of the score, the Hyv\"arinen score objective is quadratic and the estimator admits a closed-form ridge-type expression, with hyperparameters selected by an efficient leave-one-out cross-validation criterion. Third, the method enables posterior uncertainty quantification. We show that the fitted energy function can be used to recover the posterior moment generating function, extending posterior cumulant relationships for exponential-family models \citep{pericchi1993posterior}. This facilitates empirical Bayes confidence intervals through moment-based posterior discretization. We develop the framework for both Gaussian and exponential-family models and construct an analogous discrete Cf-modeling method based on Robbins' formula and a discrete Hyv\"arinen score.
Numerical studies and an application to the airway RNA-seq dataset show that the method improves point estimation and uncertainty quantification under nonlinear covariate effects, non-Gaussian latent distributions, and unequal variances \citep{himes2014rna, love2014moderated}.

The paper is organized as follows. 
Section~\ref{sec:model2} introduces conditional $f$-modeling and the conditional Tweedie formula, and develops the proposed method for Gaussian observations, including heteroscedastic extensions and posterior inference. 
Section~\ref{sec:extension} extends the framework to a general exponential family.
Sections~\ref{sec:simulation} and \ref{sec:realdata} report simulation studies and an RNA-seq application, respectively.
Concluding remarks are given in Section~\ref{sec:conclusion}.

\section{Empirical Bayes Approach via Cf-modeling}\label{sec:model2}
\subsection{Gaussian noise models}
Consider the Gaussian observation model
\begin{equation}\label{eqn:model}
y_i|\theta_i\sim N(\theta_i, \sigma^2), \qquad 
\theta_i|x_i \sim g_{x_i}(\theta_i), \qquad 
i=1,2,\ldots,n,
\end{equation}
where $g_{x_i}(\theta_i)$ represents a prior distribution of $\theta_i$ which depends on the covariates $x_i$. 
One important application is small area estimation \citep{rao2015small}, where it is generally assumed that the prior distribution of $\theta_i$ is $\theta_i \sim N(x_i^\top\beta, \tau^2)$, where $(\beta, \tau^2)$ are the hyper-parameters. Let $m(y_i|x_i)$ be the conditional marginal density function of $y_i$ defined as 
\[
m(y_i|x_i) = \int_{\theta_i} \phi(y_i;\theta_i, \sigma^2)g_{x_i}(\theta_i)d\theta_i,
\]
where $\phi(x;a,b)$ denotes the density function of a normal distribution with mean $a$ and variance $b$. 
Also, we let $s(y_i, x_i)\equiv \partial \log m(y_i|x_i)/\partial y_i$ be the conditional score function. 
In this paper, we introduce a framework for the posterior inference of $\theta_i$ based on the following fundamental theorem.

\begin{thm}\label{thm:mgf}
Under the Gaussian model (\ref{eqn:model}), the moment generating function (MGF) $M_{\theta_i}(t)$ of $\theta_i|y_i, x_i$ is expressed as  
\[
M_{\theta_i}(t)
=\exp\left(y_i t+\frac12 \sigma^2 t^2 + \int_{y_i}^{y_i+t\sigma^2} s(z,x_i)dz \right).
\]
\end{thm}

\cite{pericchi1993posterior} showed that posterior cumulants and MGFs can be represented through the marginal density in exponential-family settings when the data are exchangeable. Our work extends this perspective by showing that, in a conditional empirical Bayes framework, the conditional score function itself is sufficient to recover the posterior MGF and posterior distribution, making full posterior inference possible without estimating a normalized marginal density.

Assume the conditions of Theorem \ref{thm:mgf}. In addition, suppose that for each fixed covariate value $x_i$, the conditional marginal density $m(y_i| x_i)$ admits an analytic continuation to a complex neighborhood containing the line $\{y_i+i\sigma^2u: u\in\mathbb R\}$, and that $m(y_i| x_i)\neq 0$ on this neighborhood. Let $s(y_i,x_i)=\partial \log m(y_i| x_i)/\partial y_i$ denote the corresponding complex-valued conditional score function.

\begin{cor}\label{cor:cf}
Then the posterior characteristic function of $\theta_i| y_i,x_i$ is
\[
\varphi_{\theta_i}(t) \equiv \mathbb{E}\big[\exp(it\theta_i)|y_i,x_i\big]
=
\exp\left( ity_i-\frac{1}{2}\sigma^2t^2 + \int_{y_i}^{y_i+i\sigma^2t} s(z,x_i) dz \right).
\]
\end{cor}

Based on Theorem \ref{thm:mgf} and Corollary \ref{cor:cf}, the posterior moment generating function/characteristic function, and hence the posterior distribution, is completely determined by the conditional score function. This observation substantially broadens the role of score-based methods in empirical Bayes inference. Rather than serving solely as a tool for constructing Tweedie-type point estimators \citep{pericchi1993posterior,efron2019bayes, efron2024empirical}, as shown later in the paper, the conditional score function provides a complete characterization of the posterior distribution, enabling uncertainty quantification and interval estimation without explicit modeling of either the prior distribution or the normalized marginal density.

\subsection{Tweedie's formula}\label{sec:tweedie}
Our first application of Theorem \ref{thm:mgf} is to extend the well-known Tweedie's formula to settings with covariates. 
Under the squared error loss, it is known that the Bayes estimator of $\theta_i$ is simply the posterior mean $\mathbb{E}(\theta_i|y_i,x_i)$. We then have the following corollary:

\begin{cor}\label{cor:tweedie}
Under the Gaussian model (\ref{eqn:model}), the posterior mean of $\theta_i$ is expressed as
\[
\widehat{\theta}_i= \delta(y_i,x_i)\equiv \mathbb{E}(\theta_i | y_i,x_i)= y_i + \sigma^2 s(y_i, x_i).
\]
\end{cor}

Let $m_0(y_i|x_i)$ be the true conditional marginal density of $y_i$ given $x_i$, and let $s_0(y_i,x_i)=\partial\log m_0(y_i|x_i)/\partial y_i$ be the derivative of the true log-density. 
We then define ``oracle" Tweedie's formula as $\delta_0(y_i,x_i)=y_i+\sigma^2 s_0(y_i,x_i)$.
The difference between the two Bayes estimators (Tweedie's formulae) with true $s_0(y_i,x_i)$ and some model $s(y_i,x_i)$ can be expressed as $\sigma^2 \{s(y_i,x_i)-s_0(y_i,x_i)\}$.
Note that $\delta_0(y_i,x_i)$ is the Bayes rule under squared error loss and true density, namely, $\delta_0(y_i,x_i)=\mathbb{E}_0[\theta_i|y_i,x_i]$, where $\mathbb{E}_0$ denotes the expectation with respect to the oracle posterior with information of the true marginal density $m_0(y_i| x_i)$.
Then, we have the following excess risk decomposition:
\begin{align*}
&\mathbb{E}_0\left[ \{\theta_i-\delta(y_i,x_i)\}^2 \right] - \mathbb{E}_0\left[ \{\theta_i-\delta_0(y_i,x_i)\}^2 \right]\\
&=\mathbb{E}_0\left[ \{\delta(y_i,x_i)-\delta_0(y_i,x_i)\}^2 \right]  
- 2\mathbb{E}_0\left[ \{\theta_i-\delta_0(y_i,x_i)\} \{\delta(y_i,x_i)-\delta_0(y_i,x_i)\} \right]  \\
&=\sigma^4 \mathbb{E}_0\Big[\big\{s(y_i,x_i)-s_0(y_i,x_i)\big\}^2\Big],
\end{align*}
where the second equality follows from the fact that $\delta_0(y_i,x_i)$ is the Bayes estimator. 
The minimization of the above function is equivalent to minimizing the Fisher divergence between the true and candidate conditional marginal models.
Thus Fisher divergence directly characterizes the excess Bayes risk of Tweedie-type empirical Bayes rules.

\subsection{Energy-based modeling of the conditional density}\label{sec:model}

According to Theorem~\ref{thm:mgf}, the conditional score function determines the posterior moment generating function and hence uniquely characterizes the posterior distribution. While the posterior density can in principle be recovered through inversion of the moment generating function, such inversion is typically numerically unstable. Instead, we reconstruct the posterior distribution using a finite discrete approximation that matches the score-implied moment generating function. This yields a stable finite-dimensional representation of the posterior distribution and facilitates uncertainty quantification, confidence interval construction, and other posterior inference tasks.

For modeling the conditional density, we consider the following energy-based model: 
$$
m(y_i | x_i; \psi)=\frac{\exp\{\eta(y_i,x_i;\psi)\}}{Z(x_i; \psi)},
$$
where $\eta(y_i,x_i;\psi)$ is a flexible function parameterized by $\psi$, and 
$$
Z(x_i;\psi)=\int \exp\{\eta(y,x_i;\psi)\} dy
$$
is a normalizing constant dependent on the covariate $x_i$ and the parameter $\psi$.
Under the model, the partial derivative can be modeled as $s(y_i,x_i;\psi)=\partial \eta(y_i,x_i;\psi) / \partial y_i$.

Note that the Fisher's divergence can be written as  
\begin{align*}
\frac{1}{2}&\sigma^4 \mathbb{E}_0\Big[\big\{s(y_i,x_i;\psi)-s_0(y_i,x_i)\big\}^2\Big]\\
&=\frac{1}{2}\left\{ \sigma^4 \mathbb{E}_0[s(y_i,x_i;\psi)^2]
+2\sigma^4 \mathbb{E}_0\Big[\frac{\partial}{\partial y_i} s(y_i,x_i;\psi)\Big]
+\sigma^4 \mathbb{E}_0[s_0(y_i,x_i)^2]\right\}.
\end{align*}
Since the third term is independent of $\psi$, $\psi$ can be learned by minimizing the unbiased estimator of the first and second terms, as 
\begin{equation}\label{eq:H-score}
R(\psi)=\sigma^4\sum_{i=1}^n \left\{ 
s(y_i,x_i;\psi)^2 + 2\frac{\partial}{\partial y_i} s(y_i,x_i;\psi) 
\right\}.
\end{equation}
The objective function (\ref{eq:H-score}) is equivalent to the Hyv\"arinen score \citep{hyvarinen2005estimation}, and is the same as the Stein's unbiased risk estimate, as pointed out in \cite{ghosh2025stein}.
An attractive feature of (\ref{eq:H-score}) is that it does not depend on the normalizing constant $Z(x_i;\psi)$. 
Therefore, we can specify and estimate the score function
$s(y,x;\psi)$ directly and flexibly, without requiring an explicit form of the covariate-dependent normalizing constant. 
Provided that the resulting energy function is integrable, the fitted score induces a proper conditional density.
Importantly, the proposed empirical Bayes estimator itself only requires the fitted score function and does not require explicit evaluation of the normalizing constant. 

Motivated by this observation, we directly model the conditional score function using a flexible basis expansion. Specifically, we consider the general score model
\begin{equation}\label{eq:score-model}
s(y_i,x_i;\psi)=\sum_{j=1}^{J}\psi_j B_j(y_i,x_i;h), 
\end{equation}
where $B_j$ are basis functions that may depend on some tuning parameter $h$.
Examples include $B_k(y_i,x_i)=y_i^k$ for $k=1,\ldots,q$, $B_{q+j}(y_i,x_i)=y_ix_{ij}$, $B_{q+p+j}(y_i,x_i)=\Phi_j(x_i; h)$, and $B_{q+2p+j}(y_i,x_i)=y_i\Phi_j(x_i; h)$ for $j=1,\ldots,p$, where $\Phi_j(x_i; h)$ is a radial basis expansion $\Phi_j(x_i;h)= K\left(\|x_i-c_j\|/h\right)$, where $K(\cdot)$ is a radial basis kernel and $c_j$ and $h$ are prespecified centers and a bandwidth parameter. 

The familiar Gaussian conjugate model is a special case of (\ref{eq:score-model}). 
Suppose $\theta_i| x_i\sim N(x_i^\top\beta,\tau^2)$ and $y_i| x_i\sim N(x_i^\top\beta,\sigma^2+\tau^2),$
then the corresponding conditional score is $s(y_i,x_i;\psi)=-(y_i-x_i^\top\beta)/(\sigma^2+\tau^2)$.
Thus, the Gaussian conjugate model corresponds to a score that is linear in $y_i$ and the covariates and is naturally embedded within the general score model (\ref{eq:score-model}). 
The richer basis expansion in (\ref{eq:score-model}) allows the proposed model to accommodate nonlinear covariate effects and departures from Gaussian conjugacy.

Under the specification (\ref{eq:score-model}), we estimate $\psi$ by minimizing the penalized empirical risk
\begin{equation}\label{eq:pen-risk}
\begin{split}
L_n(\psi;h,\lambda)
&= \frac1n\sum_{i=1}^n \left[\left\{\psi^\top B(y_i,x_i)\right\}^2+2\psi^\top \frac{\partial}{\partial y_i}B(y_i,x_i)\right]+\lambda \psi^\top \psi\\
&= \frac1n\sum_{i=1}^n \left(\psi^\top Z_iZ_i^\top\psi+2W_i^\top\psi\right)+\lambda \psi^\top\psi,
\end{split}
\end{equation}
where $\psi=(\psi_1,\ldots,\psi_J)^\top$, $B(y_i,x_i)$ is the $J$-dimensional vector whose $j$th component is $B_j(y_i,x_i)$, $Z_i=B(y_i,x_i)$, and $W_i=\partial B(y_i,x_i)/\partial y_i$. 
Here $\lambda>0$ is a regularization parameter. 
The minimizer of $L_n(\psi;h,\lambda)$ is given by
\begin{equation}\label{eq:psi-hat}
\widehat{\psi}(h,\lambda)=-\left(\sum_{i=1}^n Z_iZ_i^\top+n\lambda I_J\right)^{-1}\sum_{i=1}^n W_i.
\end{equation}
To tune the hyperparameters $(h,\lambda)$, we consider the leave-one-out cross-validation criterion
\begin{equation}\label{eq:CV}
{\rm CV}(h,\lambda)=\sum_{i=1}^n \Big\{
\widehat{\psi}_{-i}(h,\lambda)^\top Z_iZ_i^\top \widehat{\psi}_{-i}(h,\lambda)+2W_i^\top \widehat{\psi}_{-i}(h,\lambda)
\Big\},
\end{equation}
where $\widehat{\psi}_{-i}(h,\lambda)$ is the minimizer of $L_n(\psi;h,\lambda)$ computed without the $i$th observation. 
While the direct computation of (\ref{eq:CV}) requires solving $n$ optimization problems, we show that ${\rm CV}(h,\lambda)$ can be evaluated in a closed form from the full-data estimator $\widehat{\psi}(h,\lambda)$.
To this end, we define the following two quantities: 
$$
M(h,\lambda)=\Big(\sum_{i=1}^n Z_iZ_i^\top+n\lambda I_J\Big)^{-1},
\qquad 
t_i(h,\lambda)=
\frac{Z_i^\top \widehat{\psi}(h,\lambda)+Z_i^\top M(h,\lambda)W_i}{1-Z_i^\top M(h,\lambda) Z_i},
$$
assuming that $M(h,\lambda)$ exists and $1-Z_i^\top M(h,\lambda) Z_i\neq 0$ for all $i=1,\ldots,n$. 
Then, the cross-validation criterion (\ref{eq:CV}) admits the following closed-form representation: 
\begin{equation}\label{eq:cv-closed}
{\rm CV}(h,\lambda) =\sum_{i=1}^n
\left[ t_i(h,\lambda)^2 + 2W_i^\top\left\{
\widehat{\psi}(h,\lambda)+M(h,\lambda)W_i+M(h,\lambda)Z_i\,t_i(h,\lambda)
\right\}\right].
\end{equation}
In particular, (\ref{eq:cv-closed}) can be evaluated without recomputing $\widehat{\psi}_{-i}(h,\lambda)$ separately for each $i$.
The derivation of (\ref{eq:cv-closed}) is provided in the Supplementary Material.
By preparing a candidate set, $\Gamma=\{(h_m,\lambda_m), \ \ m=1,\ldots,M\}$, the optimal hyperparameter can be obtained as $(h_{\rm opt}, \lambda_{\rm opt})={\rm argmin}_{(h,\lambda)\in \Gamma}{\rm CV}(h,\lambda)$.

\subsection{Posterior inference}

Based on the fitted score function, posterior inference can be carried out in a natural manner. 
First, the posterior variance of $\theta_i$ can be obtained by 
$$
{\rm Var}(\theta_i|y_i,x_i)=\sigma^2+\sigma^4 \frac{\partial}{\partial y_i}s(y_i,x_i).
$$
Therefore, once the score function is estimated, the posterior variance is also obtained directly from its derivative with respect to $y_i$. In particular, under the basis expansion in (\ref{eq:score-model}),
$$
\frac{\partial}{\partial y_i}s(y_i,x_i;\psi)=\sum_{j=1}^J \psi_j \frac{\partial}{\partial y_i}B_j(y_i,x_i),
$$
so the posterior variance can be evaluated analytically whenever the basis functions are differentiable.
In finite samples, however, the plug-in estimate $\widehat{{\rm Var}}(\theta_i|y_i,x_i)$ obtained by replacing $s(y_i,x_i;\psi)$ with $s(y_i,x_i;\widehat\psi)$ is not necessarily guaranteed to be positive because the fitted score function is estimated without imposing this constraint. 
For numerical stability, when constructing the posterior grid below, we therefore use $\widehat{{\rm Var}}_{+}(\theta_i|y_i,x_i)=\max\{\widehat{{\rm Var}}(\theta_i|y_i,x_i),\varepsilon\}$ for some small constant $\varepsilon>0$.

Next, we consider the estimation of the posterior distribution. Note that the posterior moment generating function of $\theta_i$ is
\begin{align*}
M_{\theta_i}(t)
&=\exp\left( y_i t + \frac12\sigma^2 t^2 + \int_{y_i}^{y_i+\sigma^2t} s(y,x_i;\psi)dy \right).
\end{align*}
When the score function admits the form (\ref{eq:score-model}) and the basis function $B_j(y,x;h)$ is expressed as a product form, $B_j(y,x;h)=y^{b_j}\Psi_j(x;h)$ for some constant $b_j$ and function $\Psi_j$, the integral term can be analytically evaluated and the resulting moment generating function is 
\begin{align*}
M_{\theta_i}(t)
&= \exp\left( 
y_i t + \frac12\sigma^2 t^2 + \sum_{j=1}^J \frac{\psi_j}{b_j+1}\Psi_j(x_i;h)\left\{(y_i+\sigma^2 t)^{b_j+1} - y_i^{b_j+1}\right\}
\right).
\end{align*}
It contains all information about the posterior distribution of $\theta_i$ and can produce uncertainty measures such as empirical Bayes confidence intervals.

For practical construction of empirical Bayes confidence intervals, we approximate the posterior distribution by discretization, as adopted in \cite{zhao2026nonparametric}. 
The posterior mean $\hat{\mu}_i$ and variance $\hat{v}_i$ are obtained from the first two derivatives, which are used to construct a grid for the latent signal $\theta_i$.
Specifically, for each $i$, we define grid points
$\theta_{i\ell}=\widehat{\mu}_i+\sqrt{\widehat{v}_i}z_{\ell}$ for $\ell=1,\ldots,L$, where $z_1,\ldots,z_L$ are equally spaced points on a fixed interval $[-c,c]$. 
In our implementation, we use a sufficiently large value of $c$ so that the grid covers the posterior mass. 
We then approximate the posterior distribution of $\theta_i$ by a discrete distribution
$$
\widehat{\pi}_i(\theta)
= \sum_{\ell=1}^L p_{i\ell}\delta_{\theta_{i\ell}}(\theta),
\qquad
p_{i\ell}\ge 0,
\qquad
\sum_{\ell=1}^L p_{i\ell}=1.
$$
The weights $p_{i\ell}$ are chosen so that the moment generating function of the discrete distribution matches the fitted posterior moment generating function. 
Let $t_1,\ldots,t_K$ be prespecified values around zero. 
We estimate $p_i=(p_{i1},\ldots,p_{iL})^\top$ by solving
$$
\widehat{p}_i
=\arg\min_{p_i\in\Delta_L} \sum_{k=1}^K \left[ \sum_{\ell=1}^L p_{i\ell}\exp(t_k\theta_{i\ell}) - \widehat{M}_{\theta_i}(t_k) \right]^2,
$$
where $\widehat{M}_{\theta_i}(t_k)$ is the estimated posterior moment generating function of $\theta_i$ evaluated at $t_k$ and 
$$
\Delta_L= \left\{ p_i:\ p_{i\ell}\ge 0,\ \sum_{\ell=1}^L p_{i\ell}=1\right\}.
$$
This step converts the fitted posterior moment generating function into an explicit discrete approximation of the posterior distribution. 
The posterior cumulative distribution function is then approximated by
$$
\widehat{F}_i(a)
= \sum_{\ell=1}^L \widehat{p}_{i\ell} I(\theta_{i\ell}\le a).
$$
Therefore, a $(1-\alpha)$ empirical Bayes confidence interval is constructed as $[\widehat{F}_i^{-1}(\alpha/2),\widehat{F}_i^{-1}(1-\alpha/2)]$, where
$\widehat{F}_i^{-1}(q)=\inf\{a:\widehat{F}_i(a)\ge q\}$.
This discretization approach is useful because it only requires evaluations of the fitted energy function and does not require explicit estimation of the prior distribution $g_{x_i}(\theta_i)$ or the normalizing constant of the conditional marginal density.

\subsection{Extension to unequal variances}
The proposed method can be directly extended to the case with unequal variances.
We consider the heteroscedastic model, $y_i\sim N(\theta_i, \sigma_i^2)$. 
Note that we can consider two types of loss functions for the heteroscedastic case, expressed as $L(\widehat{\theta}_i, \theta_i) = (\widehat{\theta}_i-\theta_i)^2/\sigma_i^{2a}$ for $a\in \{0,1\}$, where $a=0$ and $a=1$ correspond to the standard quadratic loss and weighted (scaled) quadratic loss, respectively. 
Under both loss functions, the Bayes estimator is
\[
\delta(y_i,x_i)\equiv 
\mathbb{E}(\theta_i|x_i, y_i) = y_i + \sigma_i^2 \frac{d}{dy_i}\log m(y_i|x_i,\sigma_i^2),
\]
while the excess risk for an empirical Bayes estimator is
\begin{align*}
& \ \ \ \ 
\frac{1}{\sigma_i^{2a}}\mathbb{E}_0\left[ \{\theta_i-\delta(y_i,x_i)\}^2 \right] 
- \frac{1}{\sigma_i^{2a}}\mathbb{E}_0\left[ \{\theta_i-\delta_0(y_i,x_i)\}^2 \right]
=\sigma_i^{4-2a} \mathbb{E}_0\Big[\big\{s(y_i,x_i)-s_0(y_i,x_i)\big\}^2\Big].
\end{align*}
Therefore, the score function can be estimated by minimizing 
$$
\sum_{i=1}^n \sigma_i^{4-2a}\left\{ 
s_i(y_i,x_i;\psi)^2 + 2\frac{\partial}{\partial y_i} s_i(y_i,x_i;\psi) 
\right\},
$$
which can be regarded as a weighted version of the original Hyv\"arinen score.
For modeling the heterogeneous score function, $s_i(y_i,x_i;\psi)$, we consider the parsimonious varying-coefficient specification
\[
\psi_j(\sigma_i^2)
=\beta_j^0+\sum_{k=1}^K\beta_kD_k(\sigma_i^2),
\qquad j=1,\ldots,J,
\]
where the variance-dependent component is shared across the $J$ score coefficients.
This specification is adopted deliberately to avoid a substantial increase in the number of parameters: allowing a separate variance-dependent component for each score coefficient would require $J(K+1)$ parameters, whereas the proposed specification requires only $J+K$ parameters.
Thus, the model allows the score function to adapt to the sampling variance while retaining a parsimonious parameterization.
The preceding arguments extend directly to this heteroscedastic setting with the corresponding unit-specific score functions.

\section{Extension to Exponential-Family Models}\label{sec:extension}
\subsection{Model and posterior moment generating functions}

The Gaussian identity in Theorem~\ref{thm:mgf} can be extended to general one-parameter exponential-family observations. Consider the conditional model
\[
p(y_i|\theta_i)
= \exp\left\{ \frac{y_i\theta_i-A(\theta_i)}{\phi} + c(y_i,\phi) \right\},
\]
where \(\theta_i\) is the canonical parameter, \(A(\cdot)\) is the cumulant function, and \(\phi>0\) is known. 
Let \(\theta_i|x_i\sim g_{x_i}\), and define the conditional marginal density $m(y_i| x_i)=\int p(y_i|\theta_i)g_{x_i}(\theta_i)d\theta_i$.
Since the carrier term \(c(y_i,\phi)\) is known, it is useful to introduce the carrier-adjusted marginal function
\[
q(y_i|x_i)
= m(y_i|x_i)\exp\{-c(y_i,\phi)\}
= \int\exp\left\{ \frac{y_i\theta-A(\theta)}{\phi} \right\} g_{x_i}(\theta)\,d\theta.
\]
Let
\[
\widetilde s(y_i,x_i)
= \frac{\partial}{\partial y_i}\log q(y_i|x_i)
= \frac{\partial}{\partial y_i}\log m(y_i|x_i) - \frac{\partial}{\partial y_i}c(y_i,\phi)
\]
denote the carrier-adjusted conditional score.
We then obtain the posterior moment generating function, as an extension of Theorem~\ref{thm:mgf}. 

\begin{thm}\label{thm:mgf-ef}
Suppose that \(q(z| x_i)\) is finite and positive on an open interval containing $\{y_i+\phi t: t\in\mathcal T\}$, where \(\mathcal T\) is a neighborhood of zero. Then, for every \(t\in\mathcal T\), the posterior moment generating function of the canonical parameter \(\theta_i| y_i,x_i\) is
\[
M_{\theta_i}(t)
=\frac{q(y_i+\phi t | x_i)}{q(y_i | x_i)}
=\exp\left\{ \int_{y_i}^{y_i+\phi t} \widetilde s(z,x_i) dz \right\}.
\]
\end{thm}

We can also derive the posterior characteristic function.

\begin{cor}\label{cor:cf-ef}
Suppose, in addition, that \(q(z|x_i)\) admits an analytic continuation to a complex neighborhood containing the line $\{y_i+i\phi u:u\in\mathbb R\}$, and that \(q(z|x_i)\neq 0\) on this neighborhood. Then the posterior characteristic function of \(\theta_i| y_i,x_i\) is
\[
\varphi_{\theta_i}(u)
=\frac{q(y_i+i\phi u| x_i)}{q(y_i| x_i)}
=\exp\left\{ \int_{y_i}^{y_i+i\phi u} \widetilde{s}(z,x_i) dz \right\}.
\]
\end{cor}

\subsection{Count response}
As a representative example of an exponential-family model, we consider the Cf-modeling framework for count responses. 
For the observed count $y_i$, we assume that $y_i|\lambda_i \sim {\rm Poisson}(\lambda_i)$, for $i=1,\ldots,n$, where $\lambda_i$ is the mean parameter and is linked to the canonical parameter through $\theta_i=\log \lambda_i$. 
Let $f(y_i|x_i)$ denote the conditional marginal probability mass function of $y_i$ given $x_i$. 
The carrier-adjusted marginal function is 
$$
q(z|x_i) = \int \exp\{z\theta-\exp(\theta)\} g_{x_i}(\theta)d\theta,
$$
which satisfies $q(y_i|x_i)=f(y_i|x_i)y!$ at non-negative integer value $y_i$. 
Theorem~\ref{thm:mgf-ef} gives the posterior moment generating function of the canonical
parameter $\theta_i$, and the result gives the posterior expectation of $\lambda_i=\exp(\theta_i)$ as 
\begin{align*}
\mathbb{E}(\lambda_i|y_i,x_i)
=M_{\theta_i}(1)
=\frac{q(y_i+1|x_i)}{q(y_i|x_i)} 
=(y_i+1)\frac{f(y_i+1|x_i)}{f(y_i|x_i)},
\end{align*}
which is known as the Robbins' formula.  
Therefore, by defining $r(y_i,x_i)=f(y_i+1|x_i)/f(y_i|x_i)$, the posterior mean of the Poisson mean parameter is expressed as $\mathbb{E}(\lambda_i|y_i,x_i)=(y_i+1)r(y_i,x_i)$.
Here $\log r(y_i,x_i)$ corresponds to the discrete difference of $\log f(y_i|x_i)$ and we define the discrete score $s(y_i,x_i)=\log r(y_i,x_i)$.
To estimate $s(y,x)$, we use a discrete Hyv\"arinen score adapted to the one-sided support $\{0,1,2,\dots\}$. 
The discrete Hyv\"arinen score $H^D(y_i,x_i;\psi)$ based on the $i$th observation $(y_i,x_i)$ is defined as
\begin{align*}
H^D(0,x_i)
&=\frac12\{s(1,x_i)+s(0,x_i)\},\\
H^D(1,x_i)
&=\frac12\{s(2,x_i)+s(1,x_i)\}
+\frac14\{s(1,x_i)+s(0,x_i)\}^2,\\
H^D(y_i,x_i)
&=\frac12\{s(y_i+1,x_i)+s(y_i,x_i)-s(y_i-1,x_i)-s(y_i-2,x_i)\}\\
&\qquad
+\frac14\{s(y_i,x_i)+s(y_i-1,x_i)\}^2,\qquad (y_i\geq 2).
\end{align*}
The empirical objective function is therefore
$$
R_{\rm dis}(\psi)=\frac{1}{n}\sum_{i=1}^n H^D(y_i,x_i; \psi)+\lambda \|\psi\|^2.
$$
This expression is obtained by writing the discrete Hyv\"arinen score in terms of the local ratio $r(y,x)$. 
In particular, for interior points $y\ge 2$, central differences are used, while for the boundary points $y=0,1$, one-sided corrections are required because the support is bounded below. 
Now we consider the basis expansion approach as in Section~\ref{sec:model}, expressed as $s(y,x;\psi)=\psi^\top B(y,x)$, where $B(y,x)$ is a vector of basis functions. 
Under this specification, $R_{\rm dis}(\psi)$ becomes a quadratic function of $\psi$, since each term in $H^D$ is either linear or quadratic in the values of the score function.
Unlike the Gaussian case, the above objective is not a SURE-type criterion. 
Instead, it depends only on the discrete score function, which allows the conditional marginal distribution to be learned without explicit normalization.

Posterior uncertainty can also be studied through the same local ratios. For instance, the posterior moments can be obtained as a function of the score function $s(y,x; \psi)$. 
Moreover, the same discrete approximation can be applied to obtain confidence intervals of $\lambda_i$.

\section{Simulation Studies}\label{sec:simulation}
\subsection{Risk evaluation and interval estimation under Gaussian response}\label{sec:sim-Gauss}

We conducted a Monte Carlo study to compare the proposed Cf-modeling method with existing empirical Bayes procedures under three different prior scenarios for the latent signal. 
In each replication, two covariates were independently generated from the uniform distribution on $[0,1]$, and the true mean structure was set to
$$
\mu(x_{i1}, x_{i2})=x_{i1}+w\left\{2\sin(2\pi x_{i1}x_{i2})+2(x_{i1}x_{i2}-1)^2\right\},
$$
where $w\in\{0,0.2,\ldots,2\}$ controls the degree of nonlinearity, which is referred to as the ``nonlinearity parameter".  
The latent signal $\theta_i$ was then generated as $\theta_i=\mu(x_{i1}, x_{i2})+v_i$.
For the true distribution of $v_i$, we adopted three cases, $v_i\sim N(0,1)$ (Gaussian), $v_i=e_i+b_i u_i$ with $e_i\sim N(0,1)$, $b_i\sim {\rm Ber}(0.05)$ and $u_i\sim U(5,10)$ (Mixture), and $v_i\sim N(0,\tau^2(x_{i1},x_{i2}))$ with $\tau(x_{i1},x_{i2})=5/[1+\exp\{-4(x_{i1}+x_{i2}-1)\}]$  (Heteroscedastic).
In all cases, the observed data were generated from $y_i=\theta_i+\ep_i$ with $\ep_i\sim N(0, \sigma_i^2)$.
We generated $\sigma_i$ from the uniform distribution on $[1, 3]$ and treated them as known values. 
We set the sample size to $n=300$ and repeated the experiment $500$ times for each value of $w$.

For each simulated dataset, we applied the proposed Cf-modeling (CF) with $J=15$ (number of basis functions) and $q=2$ (polynomial degree in $y_i$) in the score function.
For comparison, we adopted a linear empirical Bayes estimator using the prior specification, $\theta_i|x_{i1},x_{i2}\sim N(\beta_0+\beta_1x_{i1}+\beta_2x_{i2}, \gamma)$, where the parameters, $(\beta_0, \beta_1, \beta_2,\gamma)$ are estimated by the maximum likelihood method. 
Both methods provide posterior means and $90\%$ confidence intervals of $\theta_i$. 
We also consider the Gaussian deconvolution-based empirical Bayes estimator \citep{efron2016empirical} and the covariate-powered empirical Bayes method of \cite{ignatiadis2019covariate}.
Note that both methods only provide the point estimate of $\theta_i$, and that the deconvolution method cannot incorporate the covariate effect but can flexibly estimate the distribution of $y_i$.
As a benchmark, we also include the maximum likelihood estimator $y_i$ and the Wald-type $90\%$ confidence interval, $[y_i-z_{0.95}\sigma_i, y_i+z_{0.95}\sigma_i]$ with $z_{0.95}$ being the upper 5\% quantile of the standard normal distribution.

To evaluate the accuracy of point estimation, we employ the weighted mean squared errors (Weighted MSE), defined as $n^{-1}\sum_{i=1}^n (\widehat{\theta}_i-\theta_i)^2/\sigma_i^2$ for an estimator $\widehat{\theta}_i$.
For the confidence interval, we calculated the interval score \citep{gneiting2007strictly}, where the smaller value means better interval estimation in terms of coverage and length.
We also computed coverage probability, $n^{-1}\sum_{i=1}^n I(\theta_i\in {\rm CI}_i)$ and average interval length, $n^{-1}\sum_{i=1}^n |{\rm CI}_i|$, which are reported in the Supplementary Material.

Figure~\ref{fig:sim-Gauss} reports the weighted MSEs and interval scores as functions of the nonlinearity parameter $w$ under the three latent distribution scenarios. 
Overall, the proposed CF method performs favorably across all settings. 
For point estimation, it achieves the smallest or nearly smallest weighted MSE over the entire range of $w$, and its advantage becomes more pronounced as the degree of nonlinearity increases. 
This pattern is especially clear in the moderate-to-strong nonlinearity regime, where the linear empirical Bayes estimator deteriorates due to model misspecification and the deconvolution-based method suffers from its inability to incorporate covariate information.
For interval estimation, the proposed method also yields substantially smaller interval scores than the competing methods in most settings. 
As shown in the Supplementary Material, the coverage probabilities of all the methods are generally close to the nominal level, indicating that the differences in interval scores are mainly driven by differences in interval length. 
In particular, the proposed method maintains approximately nominal coverage across all scenarios, while producing substantially shorter intervals than the direct maximum likelihood method and more stable interval lengths than the linear empirical Bayes method as the degree of nonlinearity increases. 
These results suggest that the proposed score-based approach improves both point estimation and uncertainty quantification by flexibly adapting to nonlinear covariate effects and non-Gaussian latent distributions.

\begin{figure}[t]
\centering
\includegraphics[width=\textwidth]{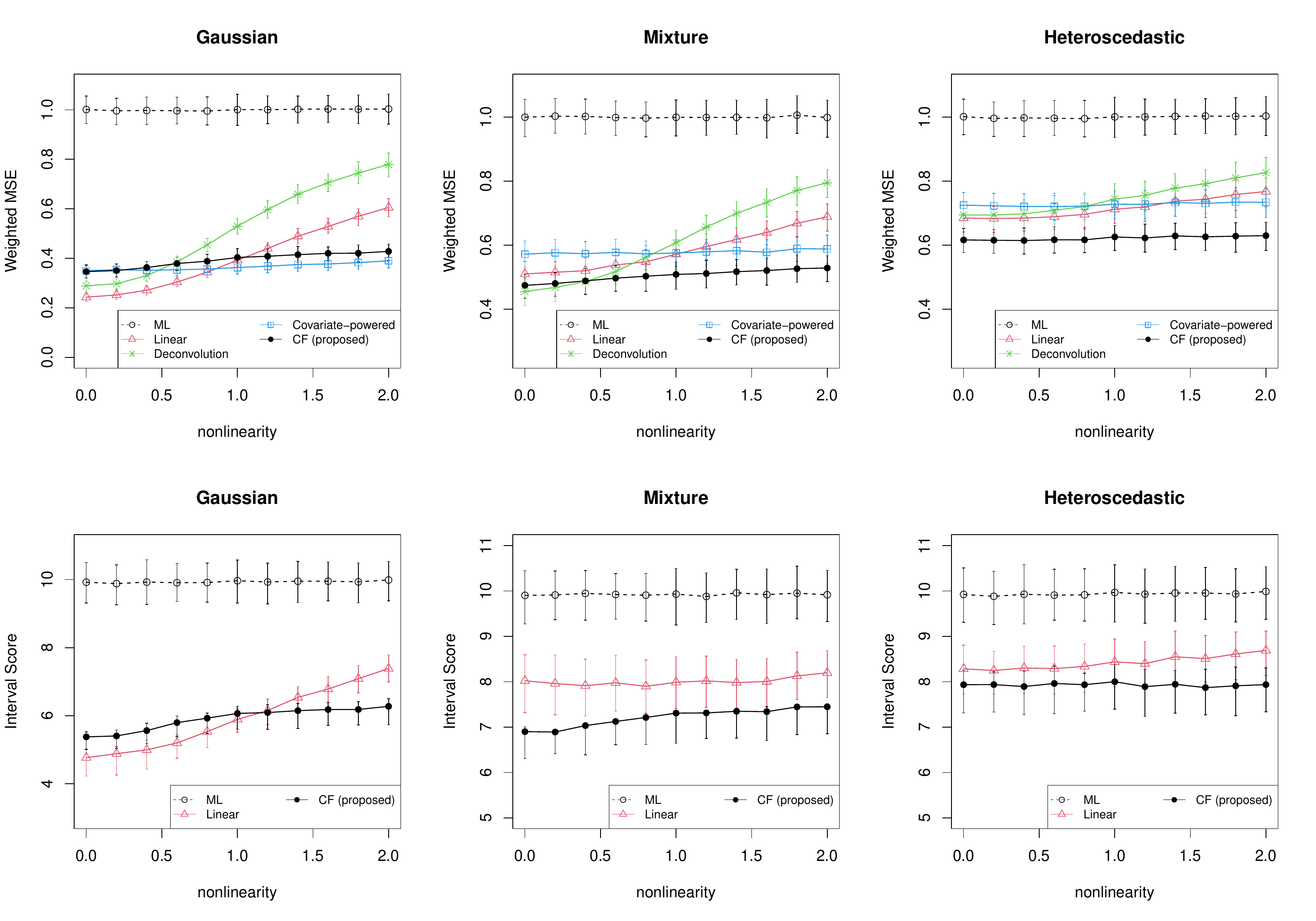}
\caption{Weighted MSE (upper) and interval score (lower) as functions of the nonlinearity parameter $w$, averaged over 500 Monte Carlo replications, under ''Gaussian'', ''Mixture'', and ''Heteroscedastic'' settings. 
The vertical bars correspond to $25\%$ and $75\%$ quantiles among 500 Monte Carlo replications. 
}
\label{fig:sim-Gauss}
\end{figure}

\subsection{Risk comparison under count response}

We next conducted a simulation study for count responses. 
The covariates were generated in the same way as in Section~\ref{sec:sim-Gauss} and the true mean structure was specified as
$$
\log\eta(x_{i1},x_{i2})=2+x_{i1}
+w\left\{2\sin(\pi x_{i1}x_{i2})+4(x_{i1}x_{i2}-1)^2-3\right\},
$$
where $w\in\{0,0.1,\ldots,1\}$ controls the complexity of the covariate effect. 
Given this mean structure, the latent intensity $\lambda_i$ was generated under three scenarios, $\lambda_i=\eta(x_{i1},x_{i2})z_i$ with $\log z_i\sim N(0, 0.2^2)$ (Log-Normal), $\lambda_i\sim {\rm Ga}(2\eta(x_{i1},x_{i2}),2)$ (Gamma) and $\lambda_i\sim{\rm Ga}((2+8z_i)\eta(x_{i1},x_{i2}),2)$ with $z_i\sim {\rm Ber}(0.05)$ (Mixture).
Then, the observed count was generated as $y_i|\lambda_i \sim {\rm Poisson}(\lambda_i)$.
We considered two sample sizes, $n=300$ and $n=600$, and repeated the experiment 500 times for each configuration.

For each simulated dataset, we applied the proposed CF method with second-order polynomial terms in $y_i$ and $J=15$ basis functions. 
As competing methods, we considered the maximum likelihood estimator $y_i$, a Poisson-Gamma empirical Bayes estimator with covariates, and a deconvolution-based empirical Bayes estimator \citep{efron2016empirical}. 
Since the present experiment focuses on point estimation, we evaluated the methods by the scaled mean squared error (scaled MSE), $n^{-1}\sum_{i=1}^n(\widehat{\lambda}_i-\lambda_i)^2/(\lambda_i+1)$.
The denominator $\lambda_i+1$ accounts for the scale of the Poisson mean while avoiding excessive instability when $\lambda_i$ is small.

\begin{figure}[t]
\centering
\includegraphics[width=\textwidth]{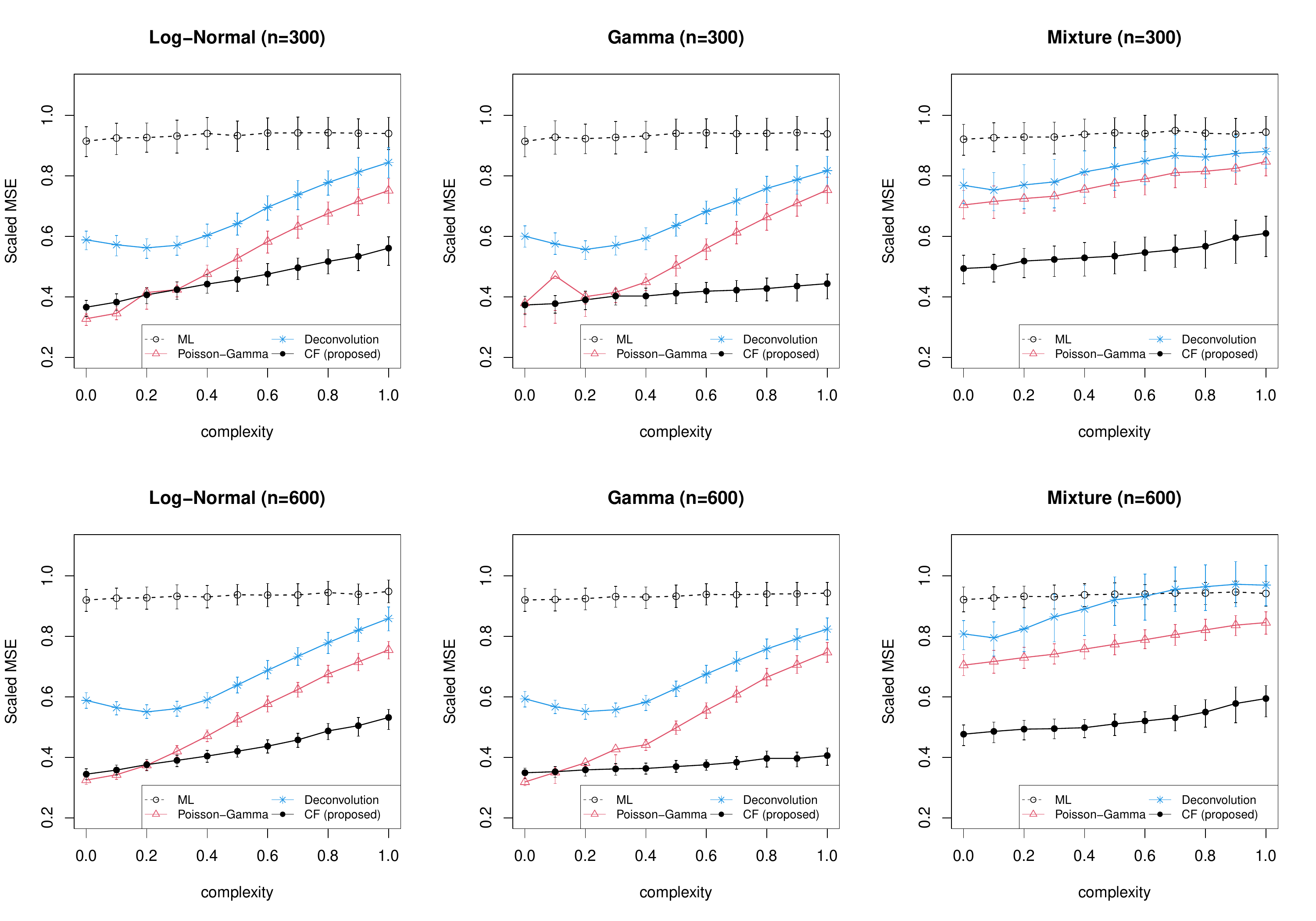}
\caption{
Scaled MSE for $n=300$ (upper) and $n=600$ (lower) as a function of the complexity parameter in the covariate effects, based on 500 Monte Carlo replications under ``Log-Normal,'' ``Gamma'' and ``Mixture'' settings.  
The vertical bars indicate the 25th and 75th percentiles
across the 500 Monte Carlo replications.
}
\label{fig:sim-Po}
\end{figure}

Figure~\ref{fig:sim-Po} reports the scaled MSEs as functions of the complexity parameter $w$ under the three latent distribution scenarios. Overall, the proposed CF method performs favorably across all settings and both sample sizes. In the log-normal and gamma scenarios, the CF method attains the smallest or nearly smallest scaled MSE over the whole range of $w$. Its advantage becomes clearer as the covariate effect becomes more complex, where the Poisson-Gamma estimator deteriorates due to its restrictive distributional structure and the deconvolution-based method suffers from ignoring covariate information. In the mixture scenario, the estimation problem is more challenging because a small fraction of observations has much larger latent means, but the proposed method remains stable and competitive. The results for $n=600$ show the same qualitative pattern as those for $n=300$, with improved accuracy, suggesting that the proposed score-based conditional modeling benefits from the larger sample size.

\section{Real Data Example: Gene-Level Differential Expression in the Airway RNA-seq Study}\label{sec:realdata}
We illustrate the proposed method using a gene-level differential expression dataset from the airway RNA-seq experiment \citep{himes2014rna}. 
The original data consist of RNA-seq read counts measured under dexamethasone-treated and untreated conditions. 
In this analysis, we focus on genes located on chromosome 17. 
For each gene, we first fitted a standard differential expression model and extracted the estimated log fold change $y_i$ and its variance $\sigma_i^2$. 
Since the estimated variances $\sigma_i^2$ exhibit substantial variation across genes, a single analysis over all genes may obscure variance-dependent shrinkage behavior. 
We therefore divided the genes into three groups according to the magnitude of $\sigma_i^2$: the lower-variance group with $\sigma_i^2<0.1$, the middle-variance group with $0.1\leq \sigma_i^2<1$, and the upper-variance group with $\sigma_i^2\geq 1$.
The sample sizes of the three groups are 753 (lower-variance), 343 (middle-variance) and 204 (upper-variance).
The proposed method and the competing empirical Bayes methods were then applied separately within each group.
In the main text, we report the results for the middle-variance group, which contains genes with moderately reliable but still non-negligibly noisy direct estimates. 
The results for the lower- and upper-variance groups are provided in the Supplementary Material. 
This stratified analysis allows us to evaluate the shrinkage behavior while avoiding the domination of the analysis by genes with extremely small or large sampling variances.

We then consider the heteroscedastic normal means model
$y_i | \theta_i,\sigma_i^2\sim N(\theta_i,\sigma_i^2)$, where $\theta_i$ denotes the true gene-specific treatment effect.
As gene-level covariates, we used two quantities, the log mean normalized expression level among control samples and the log gene length, which are denoted by $x_i$. 
These covariates are included to allow the prior distribution of the true effects to depend on gene-specific expression and structural characteristics.
We compared four empirical Bayes estimators, the proposed Cf-modeling method (denoted by CF), a linear Gaussian-Gaussian empirical Bayes estimator, the deconvolution empirical Bayes estimator, and the covariate-powered empirical Bayes estimator as used in Section~\ref{sec:sim-Gauss}.
Note that the deconvolution method does not incorporate gene-level covariates, whereas the linear estimator provides a covariate-dependent but distributionally restrictive benchmark. 
The covariate-powered method provides an additional flexible covariate-assisted benchmark.

\begin{figure}[htb!]
\centering
\includegraphics[width=\textwidth]{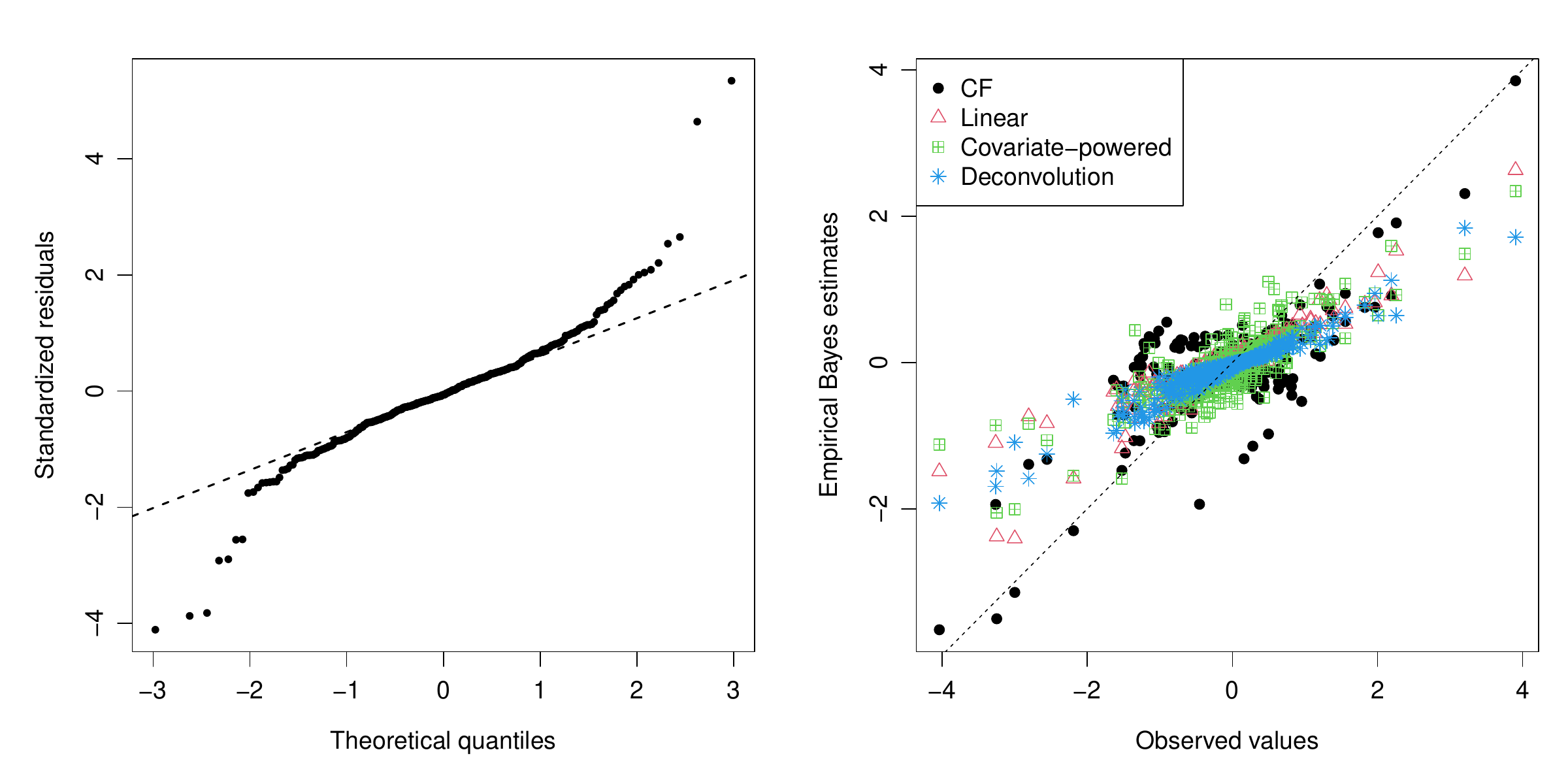}
\caption{Application to gene-level differential expression in the middle-variance group.
Left: normal Q-Q plot of the standardized residuals from the linear Gaussian-Gaussian empirical Bayes model.
Right: empirical Bayes estimates plotted against the observed values.
}
\label{fig:real-example}
\end{figure}

Figure~\ref{fig:real-example} summarizes the results. 
The left panel shows the normal Q-Q plot of the standardized residuals obtained from the linear Gaussian-Gaussian model. 
The residuals exhibit clear deviations from the Gaussian reference line, especially in the tails. 
This suggests that the Gaussian random-effects assumption is too restrictive for the present data and tends to underestimate the probability of large gene-specific effects. 
Consequently, the linear empirical Bayes estimator may overshrink large observed effects toward the fitted regression mean.
The right panel compares the empirical Bayes estimates against the observed log fold changes. 
The linear estimator shrinks the observed values strongly toward the center, which is consistent with the restrictive Gaussian random-effects assumption. 
The deconvolution estimator allows a more flexible marginal distribution, but its shrinkage pattern is essentially determined by the observed value alone because it does not use the gene-level covariates. 
In contrast, the proposed CF estimator produces a visibly more adaptive shrinkage pattern. 
For genes with similar observed log fold changes, the CF estimates can differ substantially, reflecting the use of expression level and gene length in the estimated conditional marginal score. 
This behavior indicates that the proposed method is not simply applying a univariate shrinkage rule to $y_i$, but is instead performing covariate-dependent shrinkage through the conditional marginal distribution.

We further counted the number of genes whose 90\% confidence intervals did not include zero. 
The numbers were 52 for ML, 30 for linear shrinkage and 46 for the proposed CF method. 
The smaller number for the linear method suggests that its restrictive Gaussian assumption leads to overshrinkage, especially for large observed effects. 
In contrast, CF yields substantially more intervals excluding zero than the linear method while still regularizing the noisy direct estimates.

\begin{figure}[htb!]
\centering
\includegraphics[width=\textwidth]{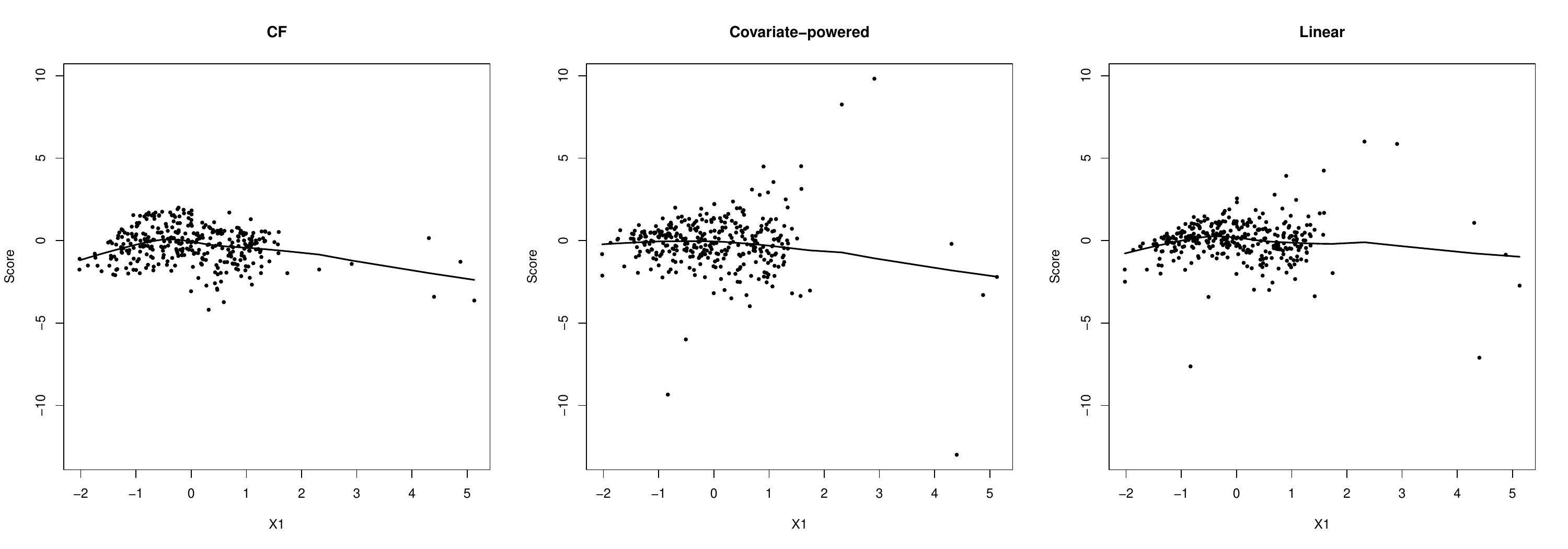}
\caption{
Estimated score values plotted against the first covariate $x_1$, the log mean normalized expression level among control samples, for the middle-variance group. 
The solid curves are locally weighted scatterplot smoothers \citep{cleveland1979robust}.
}
\label{fig:real-example-score}
\end{figure}

Figure~\ref{fig:real-example-score} further investigates the estimated shrinkage mechanism by plotting the estimated score values $(\widehat{\theta}_i-y_i)/\sigma_i^2$ against the first covariate $x_1$ (the log mean normalized expression level among control samples). 
Note that the score value is the same as the score function in the proposed CF method. 
Thus, large absolute score values can directly induce excessive shrinkage of the empirical Bayes estimates.
The proposed CF method produces a nonlinear but relatively concentrated score pattern as a function of $x_1$. 
This indicates that the estimated shrinkage rule changes flexibly with the baseline expression level while avoiding extreme score values. 
By contrast, the linear Gaussian-Gaussian estimator exhibits several outlying score values, especially for genes with moderate to large values of $x_1$. 
These extreme score values imply overly strong shrinkage adjustments and help explain the overshrinkage behavior observed in Figure~\ref{fig:real-example}. 
Therefore, the score plot provides additional evidence that the proposed CF method yields more stable and adaptive covariate-dependent shrinkage than the linear empirical Bayes benchmark.

Overall, this example highlights two advantages of the proposed approach. 
First, unlike the linear Gaussian-Gaussian empirical Bayes estimator, CF does not impose a normal random-effects distribution and can better accommodate the heavy-tailed behavior suggested by the residual diagnostic. 
Second, unlike the deconvolution-based estimator, CF incorporates gene-level covariates and can therefore adapt the shrinkage rule to systematic heterogeneity across genes. 
The results suggest that conditional score modeling provides a useful empirical Bayes framework for heteroscedastic effect estimates with covariate information.

\section{Concluding Remarks}\label{sec:conclusion}

This paper develops conditional f-modeling as a framework for empirical Bayes inference with covariate information and heterogeneous sampling distributions. The central message is that the role of the marginal score extends substantially beyond Tweedie’s formula. We show that the conditional score function determines the posterior moment generating function and, consequently, the entire posterior distribution. This observation broadens f-modeling from a tool primarily for point estimation to a framework for full posterior inference, including uncertainty quantification, without requiring explicit estimation of the latent prior distribution. More broadly, it suggests that the score function can serve as a fundamental object for empirical Bayes inference when direct modeling of the latent distribution is difficult.

Several directions merit further investigation. The basis expansion considered in this paper provides a flexible and computationally convenient model for the conditional score, but richer and more adaptive representations may be useful in problems involving high-dimensional covariates or more complex forms of heterogeneity. The reconstruction of posterior distributions from estimated scores also raises interesting questions about numerical stability, uncertainty propagation, and theoretical guarantees for posterior inference. Extensions to multivariate parameters and more general observation models would further broaden the scope of the framework. Taken together, these directions suggest that conditional score modeling may provide a useful foundation for developing flexible empirical Bayes methods beyond the classical normal means setting.

\section*{Acknowledgement}
Sugasawa's research is partially supported by JSPS KAKENHI Grant Numbers 24K21420 and 25H00546. 
Zhao's research is partially supported by the NSF grant DMS-2311216.

\vspace{0.5cm}
\bibliographystyle{chicago}
\bibliography{ref}

\newpage
\setcounter{equation}{0}
\setcounter{section}{0}
\setcounter{table}{0}
\setcounter{figure}{0}
\setcounter{page}{1}
\renewcommand{\thesection}{S\arabic{section}}
\renewcommand{\theequation}{S\arabic{equation}}
\renewcommand{\thetable}{S\arabic{table}}
\renewcommand{\thefigure}{S\arabic{figure}}

\vspace{1cm}
\begin{center}
{\LARGE
{\bf Supplementary Material for ``Beyond Tweedie's Formula: Conditional Score Modeling for Empirical Bayes Inference''}
}
\end{center}

This Supplementary Material includes technical details and additional numerical results.

\section{Derivation of the closed-form cross-validation criterion (\ref{eq:cv-closed})}

We define $S(h,\lambda)=\sum_{i=1}^n Z_iZ_i^\top +n\lambda I_J$, $U=\sum_{i=1}^n W_i$, and $u_i(h,\lambda)=Z_i^\top M(h,\lambda)Z_i$.
Then, their leave-one-out versions can be expressed as $S_{-i}(h,\lambda)=S(h,\lambda)-Z_iZ_i^\top$ and $U_{-i}=U-W_i$, and the leave-one-out estimator is $\widehat{\psi}_{-i}(h,\lambda)=-S_{-i}(h,\lambda)^{-1}U_{-i}$.
Since $S_{-i}(h,\lambda)=S(h,\lambda)-Z_iZ_i^\top$ is a rank-one modification of $S(h,\lambda)$, the Sherman--Morrison formula gives
\begin{align*}
S_{-i}(h,\lambda)^{-1}
&=
M(h,\lambda)+\frac{M(h,\lambda)Z_iZ_i^\top M(h,\lambda)}{1-Z_i^\top M(h,\lambda)Z_i}\\
&=M(h,\lambda)+\frac{M(h,\lambda)Z_iZ_i^\top M(h,\lambda)}{1-u_i(h,\lambda)}.
\end{align*}
Therefore,
\begin{align*}
\widehat{\psi}_{-i}(h,\lambda)
&=
-\left\{
M(h,\lambda)+\frac{M(h,\lambda)Z_iZ_i^\top M(h,\lambda)}{1-u_i(h,\lambda)}
\right\}(U-W_i)\\
&=
-M(h,\lambda)U+M(h,\lambda)W_i
+\frac{M(h,\lambda)Z_iZ_i^\top M(h,\lambda)(-U+W_i)}{1-u_i(h,\lambda)}.
\end{align*}
Using $-M(h,\lambda)U=\widehat{\psi}(h,\lambda)$ and $M(h,\lambda)U=-\widehat{\psi}(h,\lambda)$, we obtain
\begin{align*}
\widehat{\psi}_{-i}(h,\lambda)
&=
\widehat{\psi}(h,\lambda)+M(h,\lambda)W_i
+\frac{M(h,\lambda)Z_i\left\{Z_i^\top \widehat{\psi}(h,\lambda)+Z_i^\top M(h,\lambda)W_i\right\}}{1-u_i(h,\lambda)}.
\end{align*}
Multiplying $\widehat{\psi}_{-i}(h,\lambda)$ by $Z_i^\top$ from the left yields
\begin{align*}
Z_i^\top \widehat{\psi}_{-i}(h,\lambda)
&=
Z_i^\top \widehat{\psi}(h,\lambda)+Z_i^\top M(h,\lambda)W_i
+\frac{u_i(h,\lambda)\left\{Z_i^\top \widehat{\psi}(h,\lambda)+Z_i^\top M(h,\lambda)W_i\right\}}{1-u_i(h,\lambda)}\\
&=
\frac{Z_i^\top \widehat{\psi}(h,\lambda)+Z_i^\top M(h,\lambda)W_i}{1-u_i(h,\lambda)},
\end{align*}
which gives (\ref{eq:cv-closed}) by substituting the above quantities in (\ref{eq:CV}).

\section{Additional Simulation Results}

\begin{figure}[t]
\centering
\includegraphics[width=\textwidth]{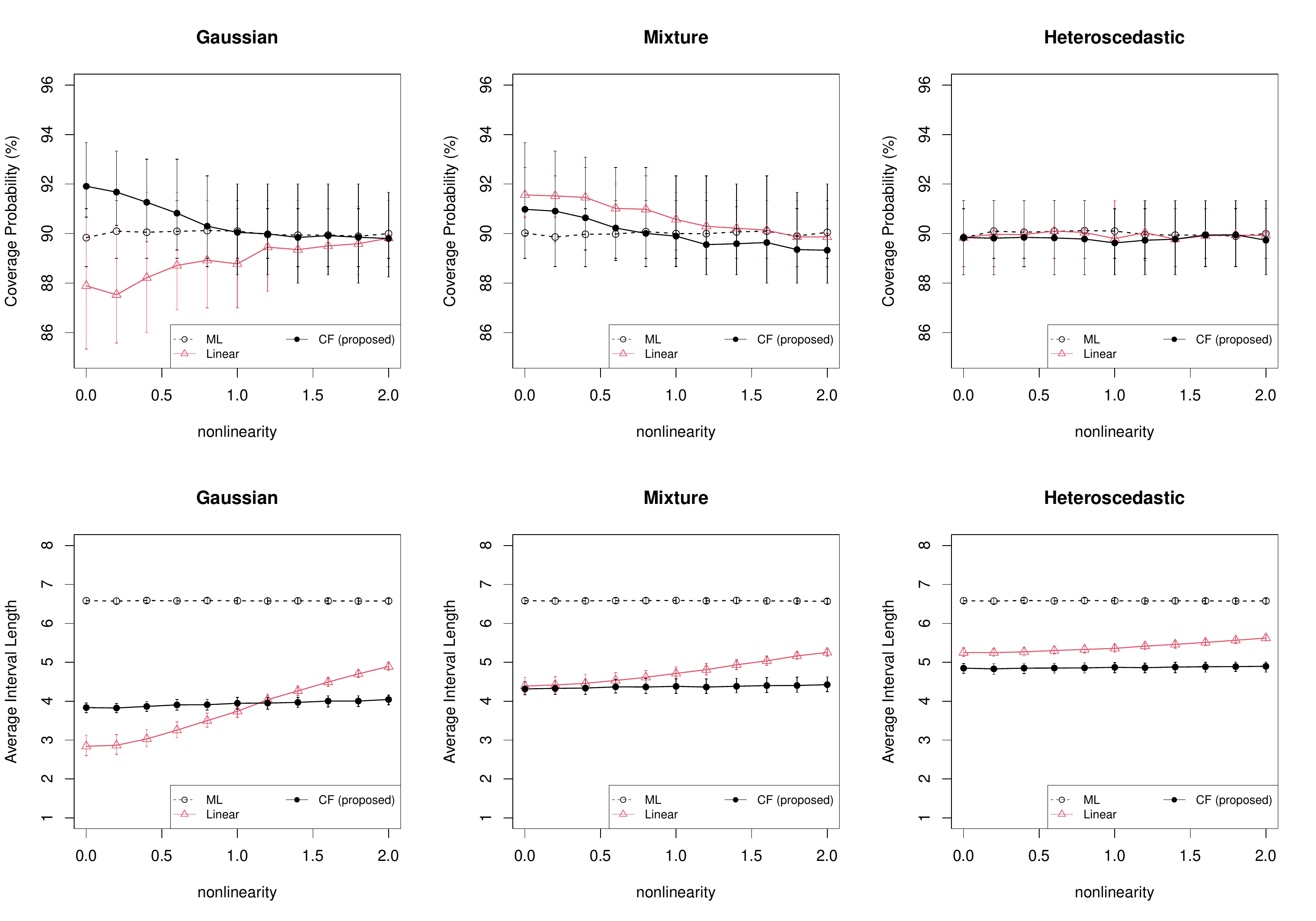}
\caption{Coverage probability (upper) and average interval length (lower) as functions of the nonlinearity parameter, averaged over 500 Monte Carlo replications, under ''Gaussian'', ''Mixture'', and ''Heteroscedastic'' settings. 
The vertical bars correspond to $25\%$ and $75\%$ quantiles among 500 Monte Carlo replications.
}
\label{fig:sim-Gauss-supp}
\end{figure}

Figure~\ref{fig:sim-Gauss-supp} reports the coverage probabilities and average interval lengths for the Gaussian-response simulation in Section~\ref{sec:sim-Gauss}.
Across the ''Gaussian'', ''Mixture'' and ''Heteroscedastic'' scenarios, the proposed CF method maintains coverage probabilities close to the nominal level over the whole range of the nonlinearity parameter.
The maximum likelihood interval also gives stable coverage, as expected from the direct Gaussian sampling model, but it produces substantially longer intervals.
The linear empirical Bayes method tends to produce shorter intervals when the nonlinearity is weak, but its interval length increases as the nonlinearity becomes stronger, reflecting the deterioration caused by misspecification of the linear Gaussian-Gaussian prior.

Overall, these additional results support the findings in the main text.
The proposed CF method achieves favorable interval scores not because of undercoverage, but because it maintains approximately nominal coverage while producing shorter intervals than the direct maximum likelihood intervals.
Compared with the linear empirical Bayes estimator, CF provides more stable uncertainty quantification under nonlinear covariate effects and non-Gaussian latent distributions.

\section{Additional Results of Real Data Example}

This section provides additional results for the lower- and upper-variance groups in the real data example of Section~\ref{sec:realdata}.
The lower-variance group is defined by $\sigma_i^2<0.1$, while the upper-variance group is defined by $\sigma_i^2\geq 1$.
The same empirical Bayes methods as in Section~\ref{sec:realdata} were applied separately to each group.

\begin{figure}[htb!]
\centering
\includegraphics[width=\textwidth]{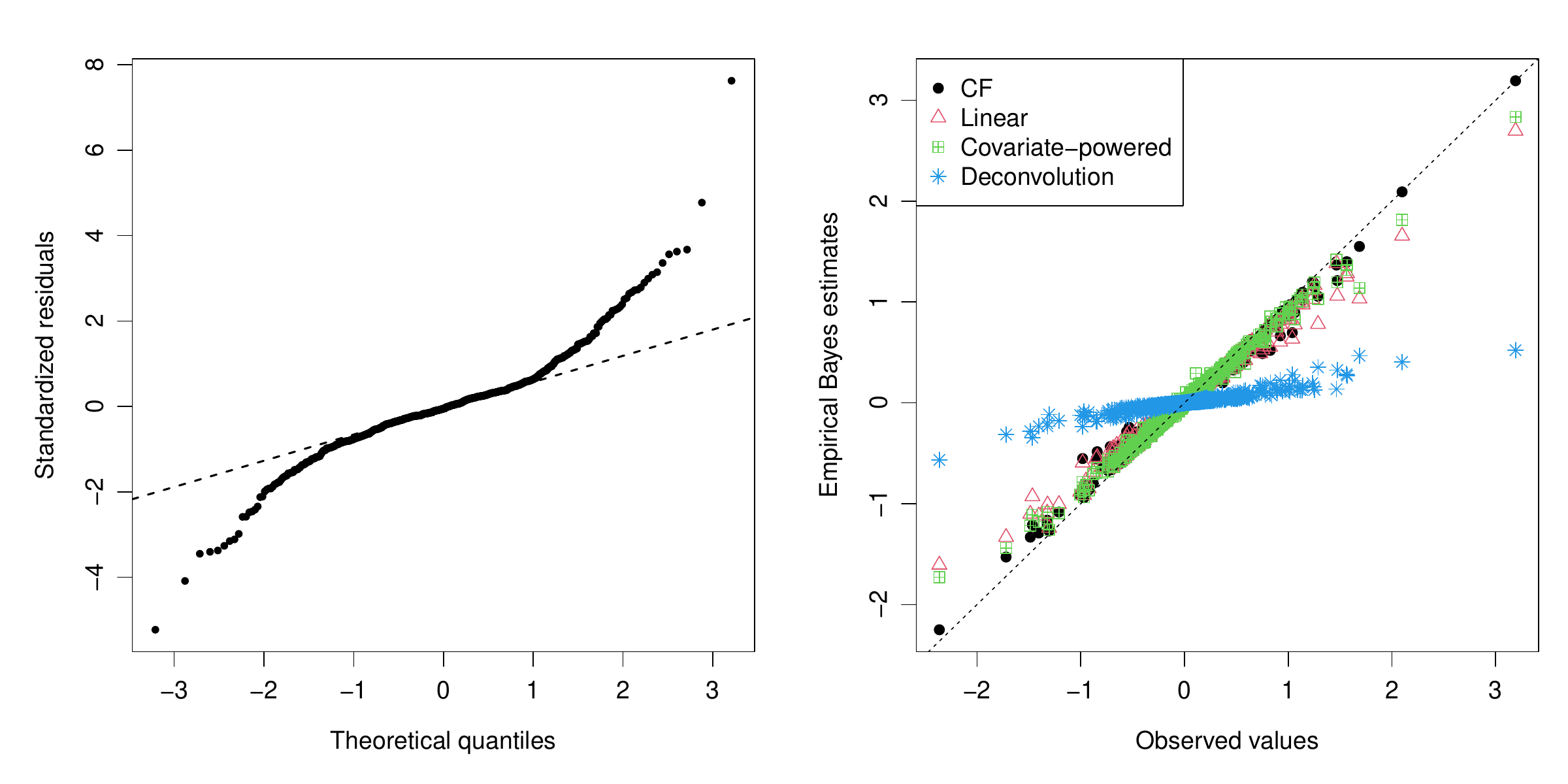}
\caption{Application to gene-level differential expression in the lower-variance group.
Left: normal Q-Q plot of the standardized residuals from the linear Gaussian-Gaussian empirical Bayes model.
Right: empirical Bayes estimates plotted against the observed values.
}
\label{fig:example-supp-lower}
\end{figure}

\begin{figure}[htb!]
\centering
\includegraphics[width=\textwidth]{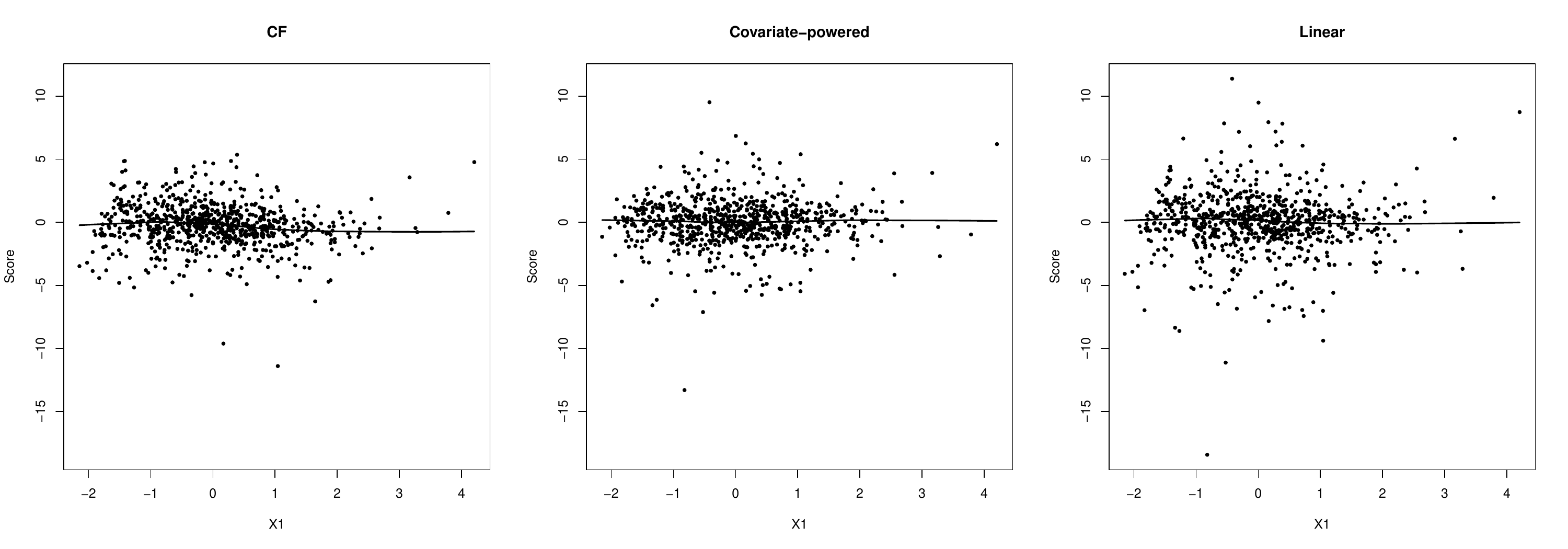}
\caption{Estimated score values plotted against the first covariate $x_1$ for the lower-variance group.
The solid curves are locally weighted scatterplot smoothers.}
\label{fig:example-supp-shrink-lower}
\end{figure}

Figures~\ref{fig:example-supp-lower} and \ref{fig:example-supp-shrink-lower} summarize the results for the lower-variance group.
The normal Q-Q plot again shows clear departures from the Gaussian reference line, especially in the tails.
This suggests that the Gaussian random-effects assumption remains restrictive even among genes with relatively small sampling variances.
In the plot of empirical Bayes estimates against the observed log fold changes, the linear estimator strongly shrinks the estimates toward the center, whereas the CF and deconvolution estimators tend to preserve large observed effects more clearly.
Since the direct estimates in this group are relatively precise, excessive shrinkage is particularly undesirable.
The score plot in Figure~\ref{fig:example-supp-shrink-lower} helps explain this behavior.
The estimated scores from CF are relatively concentrated around zero and show only a mild nonlinear trend with respect to $x_1$.
By contrast, the linear estimator produces several extreme score values, which can induce overly strong shrinkage through the heteroscedastic Tweedie's formula.
Thus, the lower-variance results also indicate that CF yields a more stable shrinkage rule than the linear Gaussian-Gaussian benchmark.

\begin{figure}[htb!]
\centering
\includegraphics[width=\textwidth]{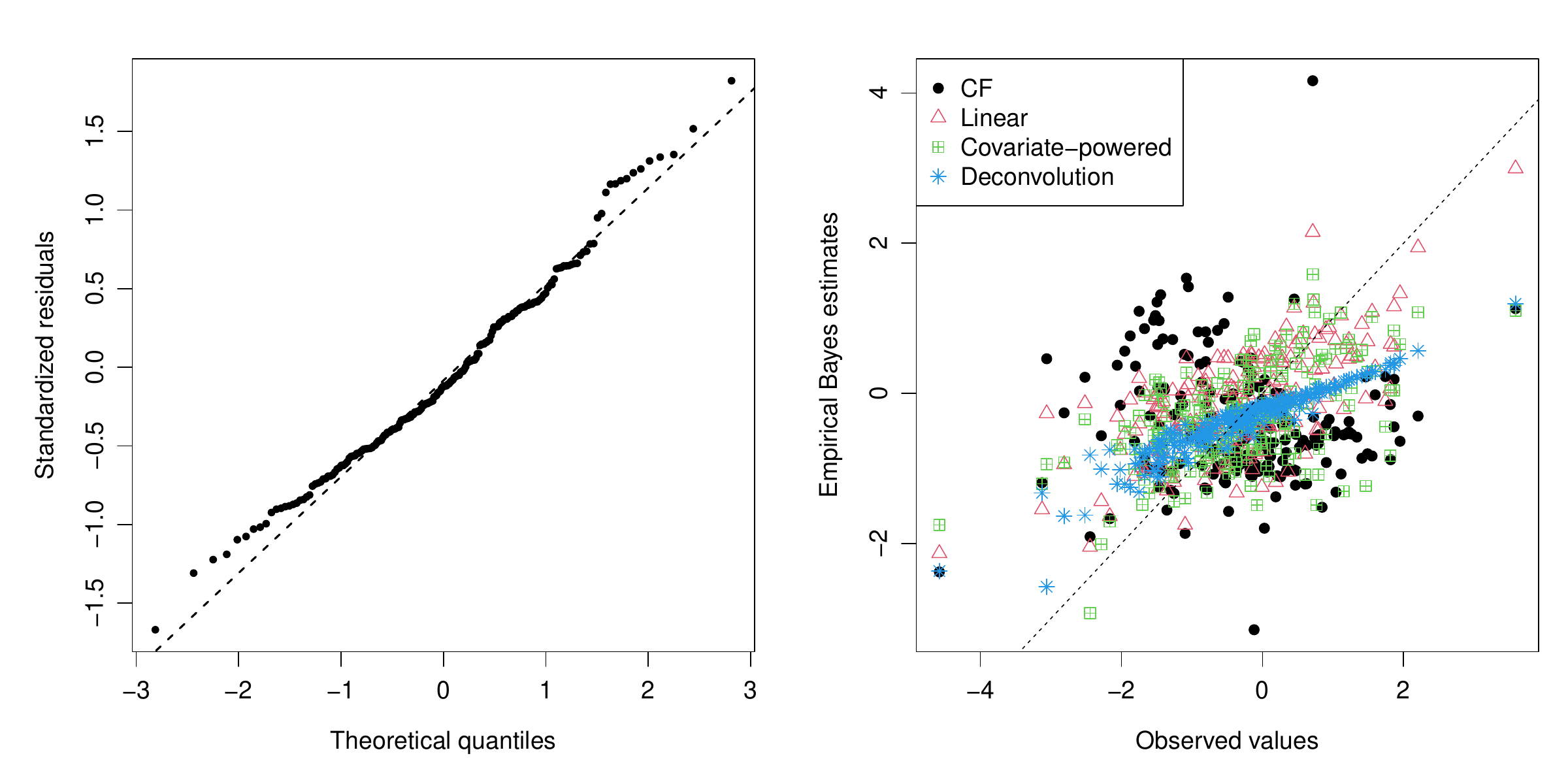}
\caption{Application to gene-level differential expression in the upper-variance group.
Left: normal Q-Q plot of the standardized residuals from the linear Gaussian-Gaussian empirical Bayes model.
Right: empirical Bayes estimates plotted against the observed values.
}
\label{fig:example-supp-upper}
\end{figure}

\begin{figure}[htb!]
\centering
\includegraphics[width=\textwidth]{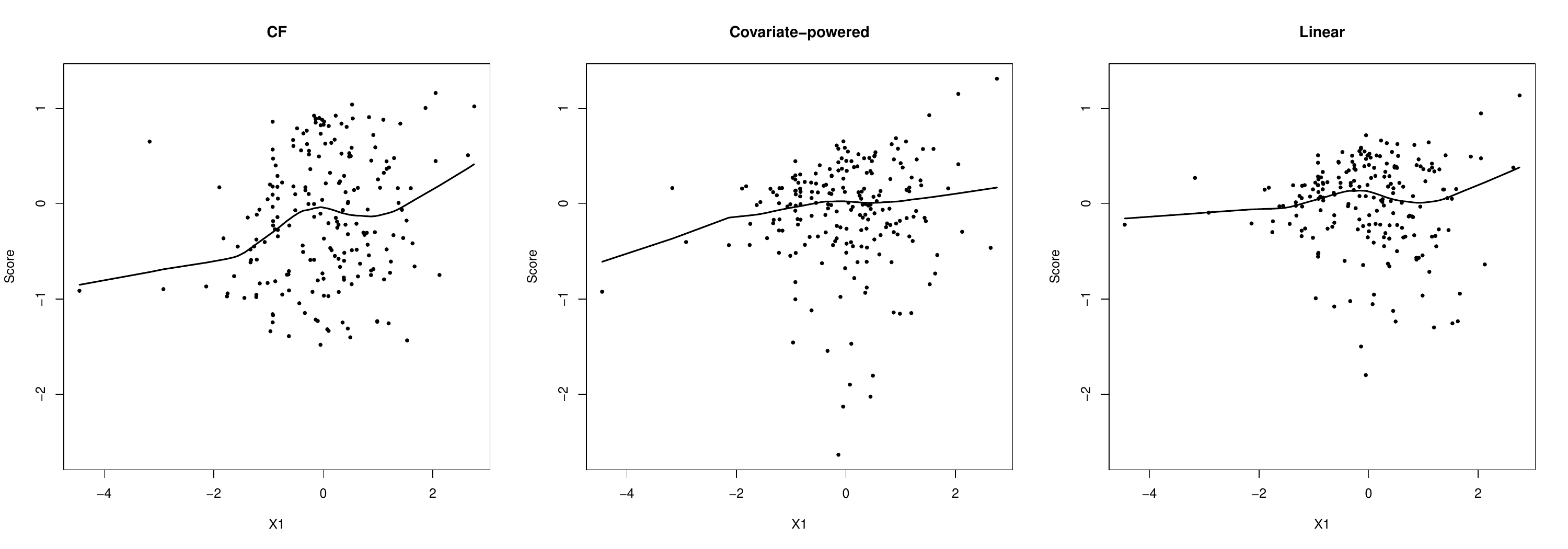}
\caption{Estimated score values plotted against the first covariate $x_1$ for the upper-variance group.
The solid curves are locally weighted scatterplot smoothers.}
\label{fig:example-supp-shrink-upper}
\end{figure}

Figures~\ref{fig:example-supp-upper} and \ref{fig:example-supp-shrink-upper} show the corresponding results for the upper-variance group.
In this group, the direct estimates are much noisier, and all empirical Bayes methods therefore apply stronger shrinkage than in the lower- and middle-variance groups.
The Q-Q plot shows that the linear Gaussian-Gaussian model is more reasonable than in the lower-variance group, although some deviations from normality remain.
The comparison of empirical Bayes estimates shows that CF still produces covariate-dependent shrinkage, while the deconvolution estimator depends only on the observed log fold change and the linear estimator is constrained by the Gaussian random-effects structure.
The score plot in Figure~\ref{fig:example-supp-shrink-upper} shows that the CF score varies nonlinearly with $x_1$, whereas the linear estimator produces a more restrictive pattern.
These results suggest that the advantage of CF is not limited to the middle-variance group reported in the main text, but is also observed across different sampling-variance regimes.

\end{document}